\documentclass[10pt,journal]{IEEEtran}

\ifCLASSOPTIONcompsoc
  \usepackage[nocompress]{cite}
\else
  \usepackage{cite}
\fi

\ifCLASSINFOpdf
 
\else
 
\fi

\usepackage{amsmath}
\usepackage{times}
\usepackage{bm}
\usepackage{amsmath}
\usepackage{amssymb,amsfonts,graphicx,float,multirow,enumitem, url}
\usepackage{bbm}
\usepackage{comment}
\usepackage{placeins}
\usepackage{graphicx,caption} 
\usepackage{blindtext} 
\usepackage{adjustbox} 

\usepackage{algorithm}
\usepackage{algorithmic}

\newtheorem{lemma}{Lemma}
\newtheorem{remark}{Remark}

\allowdisplaybreaks

\newcommand{\pr}{\textsf{pr}} 
\newcommand{\ep}{\textsf{E}} 
\newcommand{\var}{\textsf{var}} 
\newcommand{\cov}{\textsf{cov}} 

\usepackage{color} 
\definecolor{red2}{rgb}{0.7, 0, 0.1}

\usepackage{soul} 

\begin{document}

\title{Multivariate Differential Association Analysis}

\author{Hoseung~Song
        and~Michael C.~Wu 

\IEEEcompsocitemizethanks{\IEEEcompsocthanksitem Hoseung Song is with the Public Health Sciences Division at Fred Hutchinson Cancer Center, Seattle, WA 98109 USA (e-mail: hsong3@fredhutch.org). \protect \\
Michael C. Wu is with the Public Health Sciences Division at Fred Hutchinson Cancer Center, Seattle, WA 98109 USA (e-mail: mcwu@fredhutch.org).
\IEEEcompsocthanksitem }}

\markboth{Journal of \LaTeX\ Class Files,~Vol.~14, No.~8, August~2015}%
{H. Song and M.C. Wu: Multivariate differential association analysis}

\IEEEtitleabstractindextext{%
\begin{abstract}
Identifying how dependence relationships vary across different conditions plays a significant role in many scientific investigations. For example, it is important for the comparison of biological systems to see if relationships between genomic features differ between cases and controls. In this paper, we seek to evaluate whether the relationships between two sets of variables is different across two conditions.  Specifically, we assess: \textit{do two sets of high-dimensional variables have similar dependence relationships across two conditions?}. We propose a new kernel-based test to capture the differential dependence. Specifically, the new test determines whether two measures that detect dependence relationships are similar or not under two conditions. We introduce the asymptotic permutation null distribution of the test statistic and it is shown to work well under finite samples such that the test is computationally efficient, making it easily applicable to analyze large data sets. We demonstrate through numerical studies that our proposed test has high power for detecting differential linear and non-linear relationships. The proposed method is implemented in an $\texttt{R}$ package $\texttt{kerDAA}$.
\end{abstract}

}

\maketitle

\IEEEdisplaynontitleabstractindextext
\IEEEpeerreviewmaketitle


\ifCLASSOPTIONcompsoc
\IEEEraisesectionheading{\section{Introduction}\label{sec:introduction}}
\else
\section{Introduction}
\label{sec:introduction}
\fi

\IEEEPARstart{E}{valuating} dissimilarities in dependence relationships across conditions is a powerful strategy for understanding regulatory mechanisms and offers significant insights into the scientific mechanism underlying differences across groups. For instance, \cite{ezzati2016differential} studied differential association of left and right hippocampal volumes with verbal episodic and spatial memory and revealed that episodic verbal memory heavily relies on the left hippocampus, whereas spatial memory processing appears to be predominantly governed by the right hippocampus in non-demented older adults. \cite{kim2021differential} investigated the association of viral load dynamics with patient’s age and severity of COVID-19 and found that a positive correlation was observed between increased viral burden and inflammatory responses, particularly in the younger cohort and mild cases across all age groups, whereas elderly patients with critical disease exhibited minimal indications of such inflammatory responses.  These examples and countless others demonstrate that identifying disparities in co-expression/co-occurrence patterns in features (e.g. genes) between case and control groups is crucial to comprehending intricate human diseases. Such analyses, often referred to as differential co-expression (DCE) analysis, has been extensively studied \cite{bhuva2019differential, chowdhury2019differential}. DCE analysis detects pairs or sets of features that are differentially associated or regulated in different groups.

Although DCE analysis has achieved considerable success, more general application is stymied by the complex nature of the problem and the absence of robust statistical tests capable of comparing multi-dimensional patterns. Presently, DCE analyses predominantly rely on the Pearson's correlation coefficient \cite{choi2009statistical, tesson2010diffcoex, rahmatallah2014gene, mckenzie2016dgca}, which is susceptible to outliers and solely evaluates the degree of linear correlation. Pearson's correlation coefficient also only works for pairs of univariate data making it inappropriate to examine complicated associations with high-dimensional variables. Furthermore, most existing DCE analysis methods quantify the score or pattern of a pair of variables that are differentially co-expressed. This makes DCE analysis complicated and limited: testing procedures for univariate pairs are recursively applied to construct differentially co-expressed multi-variables (e.g., clusters or networks) and corresponding measures (e.g., correlation coefficients) are not effective for assessing non-linear dependence relationships or associations among high-dimensional variables. 

To bypass the limitations of extant strategies, we propose to directly perform differential association analysis by testing whether two high-dimensional variables have similar dependence relationships (beyond Pearson's correlation) or not across different conditions or groups. To measure changes in complex associations on multivariate data, we base we base our approach on the Hilbert-Schmidt Independence Criterion (HSIC) proposed by \cite{gretton2005measuring}. The HSIC is a powerful kernel-based statistic for assessing the generalized dependence between two multivariate variables and this measure makes no assumption on the distributions of the variables or the nature of the dependence.

In this work, we develop an efficient and effective kernel-based test that achieves high power in detecting changes in associations on two multivariate variables across different conditions. The main contributions of this paper are:
\begin{itemize}
    \item The new approach builds upon the HSIC and this allows the new test to work for generalized dependence relationships between two high-dimensional variables.
    \item We introduce the asymptotic permutation null distribution of the test statistic, offering an easy off-the-shelf tool for large data sets. Moreover, we apply an omnibus test with different kernels, making the new test perform well for a wide range of alternatives.
    \item The new method is implemented in an $\texttt{R}$ package $\texttt{kerDAA}$.
\end{itemize}

The organization of the paper is as follows. In the next subsection, we first introduce the permutation test, assumptions, and notations. In Section \ref{sec:hsic}, we provide an overview of the HSIC statistic. The new test with its asymptotic distribution and testing procedure is provided in Section \ref{sec:new}. Section \ref{sec:simul} examines the performance of the new tests under various simulation settings. We conclude with a brief discussion in Section \ref{sec:con}.


\subsection{Permutation tests, assumptions, and notations}

In this paper, we propose an asymptotic distribution-free test that avoids any parametric modeling assumptions. It is generally difficult to derive the true null distribution of the test statistic and this also applies to HSIC \cite{gretton2007kernel}. To overcome this, we work under the permutation null distribution, which places $1/N!$ probability on each of the $N!$ permutations of pooled observations $\{Z_{i}\}_{1,\ldots,N}$. Permutation tests are easy to implement and provide an exact control of the type I error rate for finite samples for all test statistics under the null hypothesis \cite{hoeffding1952large, lehmann2005testing}. Hence, throughout this paper, the exchangeability is assumed that the underlying distributions of samples are identical across conditions under the null hypothesis.

With no further specification, we use $\pr$, $\ep$, $\var$, and $\cov$ to denote the probability, expectation, variance, and covariance, respectively, under the permutation null distribution. In addition, we write $a_n = O(b_n)$ when $a_n$ has the same order as $b_n$ and $a_n = o(b_n)$ when $a_n$ is dominated by $b_n$ asymptotically, i.e., $\lim_{n\rightarrow\infty}(a_{n}/b_{n}) = 0$. 


\section{Hilbert-Schmidt Independence Criterion} \label{sec:hsic}

Given two random vectors $X$ and $Y$, let $f_{X}$ and $f_{Y}$ be the marginal distributions of $X$ and $Y$, respectively. We say the variables $X$ and $Y$ are statistically independent if $f_{XY} = f_{X}f_{Y}$ where $f_{XY}$ is a joint probability measure defined on $(X,Y)$.

To detect the potential associations between two sample data, the Hilbert-Schmidt independence criterion (HSIC) is widely used in many applications, such as clustering \cite{song2007dependence, niu2013iterative, he2017kernel}, time series \cite{peters2010kernel, yamada2013change, wang2021new}, and feature screening \cite{balasubramanian2013ultrahigh, freidling2021post, he2023high}. 

The HSIC was first proposed by \cite{gretton2005measuring}. The authors map the observations into a reproducing kernel Hilbert space $\mathcal{H}$ (RKHS) generated by a given kernel $k(\cdot,\cdot)$. For each point $x\in X$, there corresponds an element (feature map) $\phi(x)\in \mathcal{H}$ such that $<\phi(x), \phi(x')>_{\mathcal{H}} = k(x,x')$, where $k: X\times X \rightarrow \mathcal{R}$ is a unique positive definite kernel. With this mapping, the authors consider a cross-covariance operator between feature maps and the squared Hilbert-Schmidt norm of the cross-covariance operator, which can be expressed as
\begin{align*}
\textrm{HSIC}(X,Y) &= E_{XX'YY'}[k_{X}(X,X')k_{Y}(Y,Y')]  \\
& + E_{XX'}[k_{X}(X,X')]E_{YY'}[k_{Y}(Y,Y')] \\
& - 2E_{XY}\left[E_{X'}[k_{X}(X,X')]E_{Y'}[k_{Y}(Y,Y')]\right],
\end{align*}
where $X'$ and $Y'$ are independent copies of $X$ and $Y$, respectively. When characteristic kernels, such as the Gaussian kernel or Laplacian kernel, are used for $k_{X}$ and $k_{Y}$, it is known that $\textrm{HSIC}(f_{XY}) = 0$ if and only if $f_{XY} = f_{X}f_{Y}$. In other words, $\textrm{HSIC}$ is zero if and only if two random variables are independent.

Given $N$ pairs of observations $(X_{i},Y_{i})$ from $(X,Y) \in \mathcal{R}^{p\times q}$ $(i=1,\ldots,N)$, an empirical estimate of HSIC was also proposed as follows:
\begin{align*}
\textrm{HSIC}_{N} &= \frac{1}{N^2}\sum_{i,j}^{N}k_{X}(X_{i},X_{j})k_{Y}(Y_{i},Y_{j}) \\
& + \frac{1}{N^4}\sum_{i,j,u,v}^{N}k_{X}(X_{i},X_{j})k_{Y}(Y_{u},Y_{v}) \\
& - \frac{2}{N^3}\sum_{i,j,u}^{N}k_{X}(X_{i},X_{j})k_{Y}(Y_{i},Y_{u}).
\end{align*} 
Let $K_{X}$ and $K_{Y}$ be kernel matrices with entries $k_{X}(X_{i},X_{j})$ and $k_{Y}(Y_{i},Y_{j})$, respectively. Then, HSIC can be rewritten as
\begin{align*}
\textrm{HSIC}_{N} = \frac{trace(\tilde{K}_{X}\tilde{K}_{Y})}{N^2},
\end{align*}
where $\tilde{K}_{X} = H_{N}K_{X}H_{N}$ and $\tilde{K}_{Y} = H_{N}K_{Y}H_{N}$ are centered kernel matrices of $K_{X}$ and $K_{Y}$, respectively, and $H_{N} = I_{N} - 1_{N}1_{N}^{t}/N$ is a centering matrix with $I_{N}$ being an identity matrix of order $N$ and $1_{N}$ being a $N\times1$ vector of all ones. 


\section{New tests} \label{sec:new}

We seek to evaluate the assumption that there is underlying similar relationship between two high-dimensional sets of variables across two conditions or groups, by asking the question: \textit{do two sets of high-dimensional variables have similar dependence relationships across two conditions?}

Formally speaking, given $m$ pairs of observations $(X_{i}^{A},Y_{i}^{A}) \stackrel{iid}{\sim} f_{XY}^{A}$ on condition A where $(X_{i}^{A},Y_{i}^{A})\in \mathcal{R}^{p\times q}$ $(i=1,\ldots,m)$ and $n$ pairs of observations $(X_{i}^{B},Y_{i}^{B}) \stackrel{iid}{\sim} f_{XY}^{B}$ on condition B where $(X_{i}^{A},Y_{i}^{B})\in \mathcal{R}^{p\times q}$ $(i=1,\ldots,n)$, we concern the following hypothesis testing:
\begin{align} \label{hypo}
	H_{0}: \textrm{HSIC}^{A} = \textrm{HSIC}^{B}  \ \ vs. \ \ H_{1}: \textrm{HSIC}^{A} \ne \textrm{HSIC}^{B}.
\end{align}
The goal of a new test is to determine whether two high-dimensional variables have similar dependence relationships or not. For example, if two variables are independent in one condition, but not in the other, one of the two HSICs would be zero and the other would be greater than zero. On the other hand, if two variables are not independent in both conditions, but have different dependence relationships, two HSICs would be different.

To assess the difference of associations across conditions, the following measure is naturally considered:
\begin{align}
	T &= \textrm{HSIC}^{A} - \textrm{HSIC}^{B} \notag \\
	&= \textrm{HSIC}(X^{A}, Y^{A}) - \textrm{HSIC}(X^{B}, Y^{B}).
\end{align}
If $X$ and $Y$ are independent in both conditions, we would expect both $\textrm{HSIC}^{A}$ and $\textrm{HSIC}^{B}$ to be close to zero, then $T$ would be close to zero. If $X$ and $Y$ are not independent in both conditions, but have similar dependence relationships, we would expect $\textrm{HSIC}^{A}$ and $\textrm{HSIC}^{B}$ to be similar, then $T$ would also be close to zero. Hence, the test defined in this way is sensitive to different dependence relationships, and a value of $T$ that is far from zero would be the evidence against the null hypothesis.

Since the empirical estimate of HSIC can be written in terms of kernel matrices, we can use the following empirical measure
\begin{align*}
	T_{N} &= \textrm{HSIC}_{m}^{A} - \textrm{HSIC}_{n}^{B} \\
 &= \frac{trace(\tilde{K}_{X}^{A}\tilde{K}_{Y}^{A})}{m^2} - \frac{trace(\tilde{K}_{X}^{B}\tilde{K}_{Y}^{B})}{n^2},
\end{align*}
where $N = m + n$, $\tilde{K}_{X}^{A} = H_{m}K_{X}^{A}H_{m}$, $\tilde{K}_{Y}^{A} = H_{m}K_{Y}^{A}H_{m}$, $\tilde{K}_{X}^{B} = H_{n}K_{X}^{B}H_{n}$, and $\tilde{K}_{Y}^{B} = H_{n}K_{Y}^{B}H_{n}$. Here, we consider 
\begin{align}
	\tilde{T}_{N} &= \frac{trace\left(\left(\tilde{K}_{X}^{A}-diag(\tilde{K}_{X}^{A})\right)\left(\tilde{K}_{Y}^{A}-diag(\tilde{K}_{Y}^{A})\right)\right)}{m(m-1)} \notag \\
 &- \frac{trace\left(\left(\tilde{K}_{X}^{B}-diag(\tilde{K}_{X}^{B})\right)\left(\tilde{K}_{Y}^{B}-diag(\tilde{K}_{Y}^{B})\right)\right)}{n(n-1)},
\end{align}
where $diag(M)$ represents a matrix containing only the diagonal matrix of $M$ on its diagonal, and zero's elsewhere. Compared to $T_{N}$, $\tilde{T}_{N}$ has a tractable asymptotic permutation null distribution, while losing very little information (see below). It is also known that eliminating the diagonal terms of the cross-product matrix can reduce the bias by focusing on the dependence structure while removing the marginal effects \cite{greenacre1988correspondence, nelsen2007introduction, smilde2009matrix}.

The analytical expressions for the expectation and variance of $\tilde{T}_{N}$ and the asymptotic permutation null distribution of $\tilde{T}_{N}$ can be derived using an alternative expression for $\tilde{T}_{N}$. We first pool pairs of observations from the two conditions together and denote them by $(X_{i}^{U},Y_{i}^{U}) = (X^{A},Y^{A})\cup (X^{B},Y^{B})$ $(i=1,\ldots,N)$. Let $K_{X}^{U}$ and $K_{Y}^{U}$ be kernel matrices for the pooled observations $X_{i}^{U}$ and $Y_{i}^{U}$, respectively $(i=1,\ldots,N)$. Let 
\begin{align*}
	H = \begin{pmatrix}
		H_{m} & \textbf{0}_{m\times n} \\
		\textbf{0}_{n\times m} & H_{n}
	\end{pmatrix},
\end{align*}
where $\textbf{0}_{m\times n}$ and $\textbf{0}_{n\times m}$ are $m\times n$ and $n\times m$ matrices of all zeros, respectively. For centered matrices $\tilde{K}_{X}^{U} = HK_{X}^{U}H$ and $\tilde{K}_{Y}^{U} = HK_{Y}^{U}H$, let $\tilde{K}^{U}$ be a kernel matrix with entries $\tilde{k}_{X}^{U}(X_{i},X_{j})\tilde{k}_{Y}^{U}(Y_{i},Y_{j})$ $(i,j=1,\ldots,N)$, namely the element-wise products between $\tilde{K}_{X}^{U}$ and $\tilde{K}_{Y}^{U}$, with diagonal elements set to zero. We denote the $(i,j)$-th element of $\tilde{K}^{U}$ by $\tilde{k}_{ij,U}$. Then, $\tilde{T}_{N}$ can be rewritten as:
\begin{align} \label{stat}
    \tilde{T}_{N} = \frac{1}{m(m-1)}\sum_{i,j=1}^{m}\tilde{k}_{ij,U} - \frac{1}{n(n-1)}\sum_{i,j=m+1}^{N}\tilde{k}_{ij,U}.
\end{align}

The analytical expressions for the expectation and variance of $\tilde{T}_{N}$ and its asymptotic permutation null distribution can be obtained in a similar way to that in \cite{song2020generalized} and are provided in Lemma 1 and 2.

\begin{lemma} \label{lemma1}
    Under the permutation null distribution, we have
    \begin{align*}
        \ep(\tilde{T}_{N}) &= 0, \\
        \var(\tilde{T}_{N}) &= \frac{2A f_1(m)+4B f_2(m)+C f_3(m)}{m^2(m-1)^2} \\
        &+ \frac{2A f_1(n)+4B f_2(n)+C f_3(n)}{n^2(n-1)^2} \\
        &- \frac{2C}{N(N-1)(N-2)(N-3)},
    \end{align*}
    where 
    \begin{align*}
	f_1(x) & = \frac{x(x-1)}{N(N-1)}, \ \ f_2(x)  = \frac{x(x-1)(x-2)}{N(N-1)(N-2)}, \\ f_3(x) &= \frac{x(x-1)(x-2)(x-3)}{N(N-1)(N-2)(N-3)}, \\
	A &= \sum_{i,j=1}^{N}\tilde{k}^2_{ij,U}, \ \ B = \sum_{i,j,r=1}^{N}\tilde{k}_{ij,U}\tilde{k}_{ir,U}, \\ 
	C &= \sum_{i,j,r,s=1}^{N}\tilde{k}_{ij,U}\tilde{k}_{rs,U}.
	\end{align*}
\end{lemma}

We then define the test statistic as
\begin{align}
    Z_{\tilde{T}} = \frac{\tilde{T}_{N} - \ep(\tilde{T}_{N})}{\sqrt{\var(\tilde{T}_{N})}}.
\end{align}
\begin{lemma} \label{lemma2}
    Let $\bar{k} = \sum_{i,j=1}^{N}\tilde{k}_{ij, U}/(N^2-N)$ and $\bar{k}_{i\cdot} = \sum_{j=1,j\ne i}^{N}(\tilde{k}_{ij,U} - \bar{k})$ for $i=1,\ldots,N$. Under the permutation null distribution, as $N\rightarrow\infty$, $m/N \rightarrow p \in (0,1)$, if
    $\sum_{i=1}^{N}|\bar{k}_{i\cdot}|^\epsilon = o(\{\sum_{i=1}^{N}\bar{k}_{i\cdot}^2\}^{\epsilon/2})$ for all integers $\epsilon>2$, \begin{align*}
        Z_{\tilde{T}}  \stackrel{\mathcal{D}}{\rightarrow} \mathcal{N}(0,1).
    \end{align*}
\end{lemma}

\begin{remark}
	The condition for Lemma \ref{lemma2} can be satisfied when $|\bar{k}_{i\cdot}| = O(N^{\delta})$ for a constant $\delta$, $\forall i$. When there is no big outlier in the data, it is not hard to have this condition satisfied.
\end{remark}

\begin{figure}[h]
	\centering
	\includegraphics[width=\columnwidth]{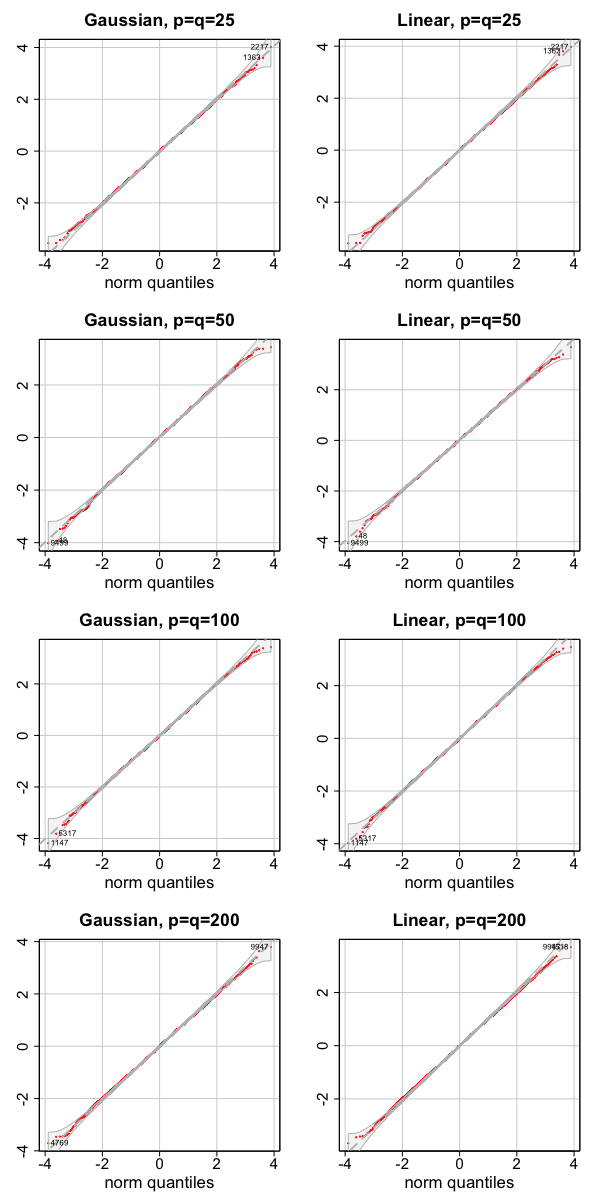}
	\caption{\label{qq}Quantile-quantile plots of $Z_{\tilde{T}}$ using the Gaussian kernel and linear kernel when $N=200$.}
\end{figure}

Figure \ref{qq} shows the normal quantile-quantile plots for $Z_{\tilde{T}}$ from 10,000 permutations under different choices of kernels and $p,q$ for Gaussian data with $N = 200$. We see that, when $N$ is in the hundreds, the permutation distributions can already be well approximated by the standard normal distribution for $Z_{\tilde{T}}$.

To assess the dependence relationship, we use the Gaussian kernel and linear kernel as each kernel is suitable for a different type of dependence relationship \cite{gretton2007kernel, liu2021kernel, liu2023kernel}. To accommodate both effects based on the Gaussian kernel and linear kernel, we apply the Cauchy combination test to obtain the omnibus $p$-value \cite{liu2020cauchy}. The detailed testing procedure is summarized in Algorithm \ref{algo}.
\begin{algorithm}[h]
	\caption{Testing procedure} \label{algo}
	\begin{algorithmic}[1]
		\REQUIRE Observations $\{(X_{i}^{U},Y_{i}^{U})\}_{i=1,\dots,N}$ and the significance level $\alpha$.
		\ENSURE Reject the null hypothesis $H_{0}$ if $p$-value $p \le \alpha$.
		\STATE Compute $Z_{\tilde{T}}$ using the Gaussian kernel and linear kernel by Lemma \ref{lemma1}.
		\STATE Calculate $p$-values of $Z_{\tilde{T}}$ by the standard normal distribution based on Lemma \ref{lemma2} using the Gaussian kernel and linear kernel ($p_{G}$ and $p_{L}$).
		\STATE Obtain the omnibus $p$-value:
        \begin{align}
            p &= \frac{\tan\{(0.5-\min(p_{G},0.99))\pi\}}{2} \notag \\
            &\ \ \ \ +\frac{\tan\{(0.5-\min(p_{L},0.99))\pi\}}{2}. \notag
        \end{align}
	\end{algorithmic}
\end{algorithm}


\section{Experiments} \label{sec:simul}


\subsection{Simulation studies}

In this section, we examine the performance of the new test under various simulation examples. We compare the new test (NEW) with the existing correlation-based test (dCoxS) proposed by \cite{cho2009identifying}. The authors utilize the Pearson's correlation coefficient between two pairwise Euclidean distances of samples. dCoxS compares the Pearson's correlation coefficient between two conditions and calculates a z-score by applying the Fisher’s transformation to the Pearson’s correlation coefficients. Similarly, many methods are based on the correlation coefficient, which focuses on the linear dependence relationships \cite{choi2009statistical, tesson2010diffcoex, amar2013dissection, siska2016discordant}. Here, we apply the permutation test to dCoxS for fair comparison.

We first consider the following settings (Setting 1): given $\{(X_{i}^{A},Y_{i}^{A})\}  \stackrel{iid}{\sim}  f_{XY}^{A}$ vs. $\{(X_{i}^{B},Y_{i}^{B})\}  \stackrel{iid}{\sim}  f_{XY}^{B}$,
\begin{itemize}
    \item Multivariate normal: $f_{XY}^{A} \sim N_{100}(\textbf{0}_{100},I_{100})$ and $f_{XY}^{B} \sim N_{100}(\textbf{0}_{100},\Sigma)$.
    \item Multivariate log-normal: $f_{XY}^{A} \sim \exp\left(N_{100}(\textbf{0}_{100},I_{100})\right)$ and $f_{XY}^{B} \sim \exp\left(N_{100}(\textbf{0}_{100},\Sigma)\right)$.
\end{itemize}
We use $m=n=100$, $p=q=50$, and $\Sigma = \rho^{|i-j|}$. $\textbf{0}_{d}$ represents $d$ dimensional vector of zeros. When $\rho = 0$, both $\{(X_{i}^{A},Y_{i}^{A})\}$ and $\{(X_{i}^{B},Y_{i}^{B})\}$ are independent, and we would expect the new test does not reject the null hypothesis. On the other hand, when $\rho > 0$, $\{(X_{i}^{B},Y_{i}^{B})\}$ are not independent each other, while $\{(X_{i}^{A},Y_{i}^{A})\}$ are still independent each other, we thus expect the new test rejects the null hypothesis. We simulate 1000 data sets to estimate the power of the tests and the significance level is set to be 0.05 for all tests. 

\begin{table}[h!]
	\caption{\label{type1}Empirical size of the tests at 0.05 significance level}
	\centering
	\begin{tabular}{c|c|c}
		\hline
		& NEW & dCoxS \\ \hline
		Normal & 0.047 & 0.038  \\ \hline
		Log-normal & 0.045 & 0.047  \\ \hline
	\end{tabular}
\end{table}

Table \ref{type1} shows the empirical size of the tests at 0.05 significance level for the multivariate normal and log-normal data. We see that both the new test and dCoxS control the type I error rate well.

\begin{figure}[h!]
	\centering
	\includegraphics[width=\columnwidth]{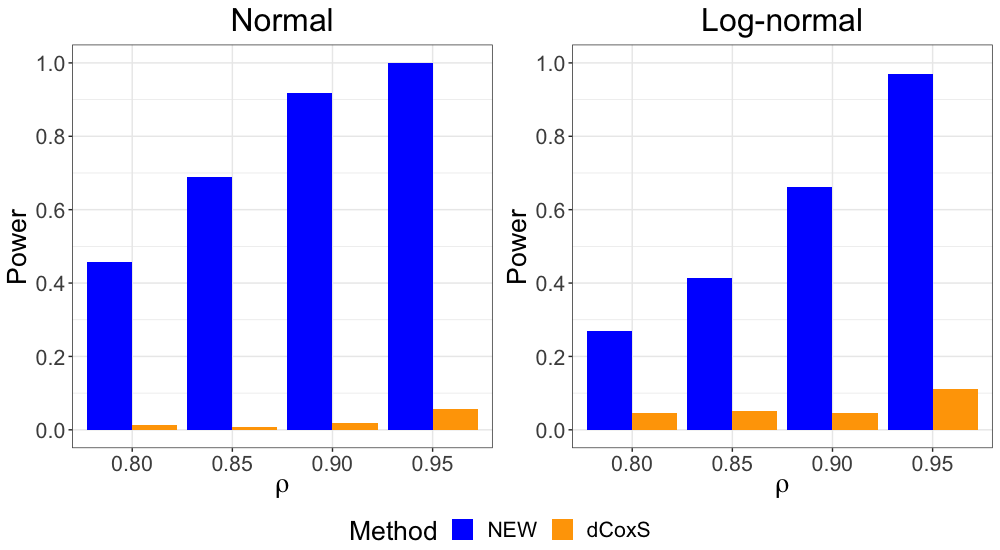}
	\caption{\label{power1}Estimated power of the tests under different $\rho$ when $m=n=100$ and $p=q=50$.}
\end{figure}

Figure \ref{power1} shows the estimated power of the tests under different $\rho$ when $m=n=100$ and $p=q=50$. Since the dependence relationship between $X_{i}$ and $Y_{i}$ gets stronger as $\rho$ increase, we would expect the power of tests increases as $\rho$ increases. We see that the performance of the new test indeed increases as $\rho$ increases, but dCoxS performs poorly.

We also consider different types of settings (Setting 2):
\begin{itemize}
    \item Case 1: $\{(X_{i}^{A},Y_{i}^{A})\} \stackrel{iid}{\sim}  N(\textbf{0}_{100}, \Sigma_{1})$ vs. $\{(X_{i}^{B},Y_{i}^{B})\} \stackrel{iid}{\sim}  N(\textbf{0}_{100}, \Sigma_{2})$, where $\Sigma_{1} = 0.4^{|i-j|}$ and $\Sigma_{2} = (0.4+\rho)^{|i-j|}$.
    \item Case 2: $\{(X_{i}^{A},Y_{i}^{A})\} \stackrel{iid}{\sim}  \exp(N(\textbf{0}_{100}, \Sigma_{1}))$ vs. $\{(X_{i}^{B},Y_{i}^{B})\} \stackrel{iid}{\sim}  \exp(N(\textbf{0}_{100}, \Sigma_{2}))$, where $\Sigma_{1} = 0.4^{|i-j|}$ and $\Sigma_{2} = (0.4+\rho)^{|i-j|}$.
    \item Case 3: $\{X_{i}^{A}\} \stackrel{iid}{\sim}  N_{100}(\textbf{0}_{100},I_{100})$ and $\{Y_{i}^{A}\} = \sin(2\pi X_{i,1:50}^{A}/3)$ vs. $\{X_{i}^{B}\} \stackrel{iid}{\sim}  N_{100}(\textbf{0}_{100},I_{100})$ and $\{Y_{i}^{B}\} = \sin((2+\rho)\pi X_{i,1:50}^{B}/3)$, where $X_{i,1:50}^{A}$ and $X_{i,1:50}^{B}$ are the first 50 variables of $X_{i}^{A}$ and $X_{i}^{B}$, respectively.
    \item Case 4: $\{X_{i}^{U}\} \stackrel{iid}{\sim}  \log|N_{100}(\textbf{0}_{100},I_{100})|$ and $\{Y_{i}^{U}\} \stackrel{iid}{\sim}  \sin(N_{100}(\textbf{0}_{100},I_{100}))$, where the first 10 variables of  $\{Y_{i}^{A}\}$ depend on the first variable of $\{X_{i}^{A}\}$ and the first $10+\rho$ variables of  $\{Y_{i}^{B}\}$ depend on the first variable of $\{X_{i}^{B}\}$.
\end{itemize}
We use $m=n=100$ and $p=q=50$. Setting 2 is somewhat different from Setting 1: in Setting 1, two variables are independent in one condition, but not independent in the other, while in Setting 2, two variables are not independent in both conditions, but dependence relationships could be different. In Setting 2, when $\rho=0$, two variables have the same dependence relationships in both conditions, so we would expect the new test does not reject the null hypothesis. On the other hand, when $\rho > 0$, two variables have different dependence relationships, so we would expect the new test rejects the null hypothesis. We simulate 1000 data sets to estimate the power of the tests and the significance level is set to be 0.05 for all tests. 

\begin{table}[h!]
	\caption{\label{type2}Empirical size of the tests at 0.05 significance level}
	\centering
	\begin{tabular}{c|c|c}
		\hline
		& NEW & dCoxS \\ \hline
		Case 1 & 0.048 & 0.040  \\ \hline
		Case 2 & 0.054 & 0.043  \\ \hline
        Case 3 & 0.046 & 0.047  \\ \hline
        Case 4 & 0.052 & 0.064  \\ \hline
	\end{tabular}
\end{table}

The empirical size of the tests at 0.05 significance level for Setting 2 is presented in Table \ref{type2}. We see that both the new test and dCoxS control the type I error rate well in all cases.

\begin{figure}[h!]
	\centering
	\includegraphics[width=\columnwidth]{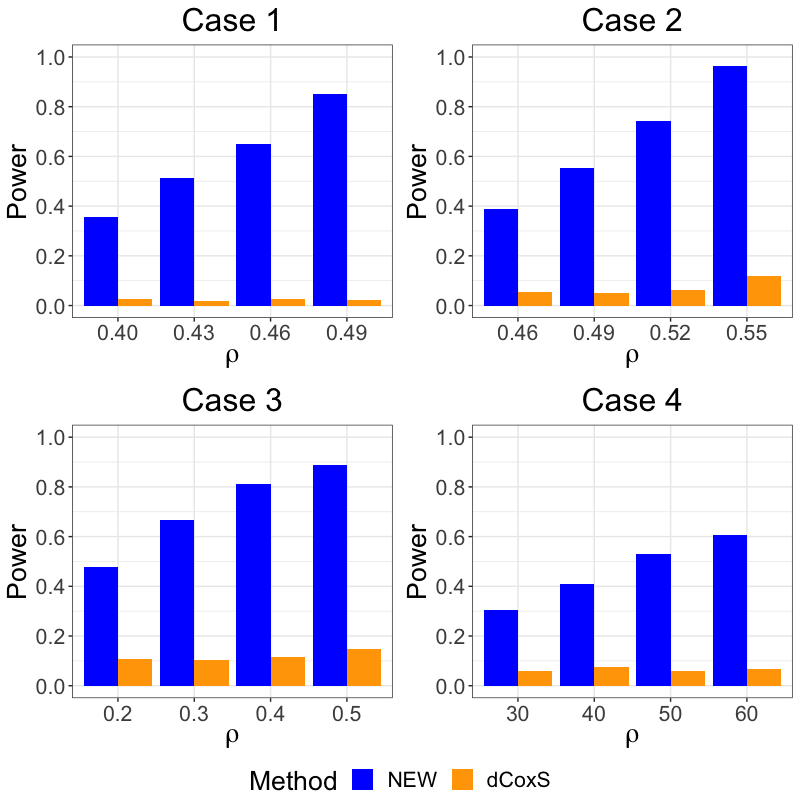}
	\caption{\label{power2}Estimated power of the tests under different $\rho$ when $m=n=100$ and $p=q=50$.}
\end{figure}

Figure \ref{power2} shows the performance of the tests under different $\rho$ for all cases. Similar to Setting 1, a large value of $\rho$ indicates strong dependence relationships. Following this, the power of the new test increases as $\rho$ increases and the new test exhibits high power. However, dCoxS does not capture the difference in dependence relationships.


\subsection{Application to real data}

We apply the new test to the data from Menopause Studies - Finding Lasting Answers for Symptoms and Health (MsFLASH) Vaginal Health Trial \cite{mitchell2018efficacy}. The MsFLASH trial was a randomized clinical trial to evaluate the treatment effect of vaginal estradiol vs. placebo on vaginal discomfort in postmenopausal women. In an endeavor to explore the mechanisms underlying postmenopausal vaginal discomfort, a longitudinal investigation was conducted to study the vaginal microbiota and vaginal fluid metabolites of 141 participants. The vaginal microbiome profiles include abundance data for 381 taxa and the metabolome profiles include abundance data for 171 metabolites.

We applied the proposed test to assess differential dependency between metabolites and the overall vaginal microbiome compositions across the placebo (45 participants) and estradiol treatment arms (50 participants). Specifically, we applied the new test to assess the differential dependence relationships at baseline and week 4. We obtained $p$-values 0.937 and 0.038 for participants at baseline and week 4, respectively. This result shows that the dependence relationship between metabolites and vaginal microbiome compositions was not different by arm at baseline.  This makes sense as none of the women had undergone treatment yet.  However, after four weeks of treatment, the dependency between metabolites and microbiome composition was significantly different ($\alpha = 0.05$). Hence, this result shows that the association between metabolites and vaginal microbiome compositions is perturbed over time. 

\begin{table}[h]
	\caption{\label{real:gau}HSIC values using the Gaussian kernel}
	\centering
	\begin{tabular}{c|c|c}
		\hline
		& HSIC: Placebo & HSIC: Estradiol \\ \hline
		Baseline & 0.0051 &  0.0046 \\ \hline
		Week 4 & 0.0056 & 0.0047 \\ \hline
	\end{tabular}
\end{table}

\begin{table}[h]
	\caption{\label{real:lin}HSIC values using the linear kernel}
	\centering
	\begin{tabular}{c|c|c}
		\hline
		& HSIC: Placebo & HSIC: Estradiol \\ \hline
		Baseline & 6883.3 &  5474.8 \\ \hline
		Week 4 & 3762.3 & 7127.6 \\ \hline
	\end{tabular}
\end{table}

We further investigated HSIC values for treatment and control groups over time. Table \ref{real:gau} and \ref{real:lin} present HSIC values for the control (placebo) and treatment (Estradiol) groups over time (baseline vs. week 4) using the Gaussian kernel and linear kernel, respectively. We see that, though HSIC values using the Gaussian kernel are difficult to distinguish from each other, the HSIC value using the linear kernel for the treatment group is significantly larger than the value for the control group at week 4. This result indicates that the new test detects the perturbation in the dependence relationship between the control and treatment groups over time.


\section{Conclusion and discussion} \label{sec:con}

In this paper, we proposed a new kernel-based test for evaluating whether pairs of high-dimensional variables have similar dependence relationships or not across two conditions. Using the previously developed HSIC statistic for our test statistic, the new test works for high-dimensional data and performs well for assessing complicated dependence relationships. The asymptotic distribution of the test statistic facilitates its application to large data sets and the omnibus test enables the proposed test to work for a wide range of alternatives. The new test often exhibits superior power without particular model assumptions or specifications.

This paper mainly handles the generalized dependence relationships caused by multivariate data. Compared to the existing DCE analysis that quantifies differential correlated patterns across conditions, the focus of this paper is to study whether the new test captures the change in dependence relationships across different conditions. Hence, the new test does not seek to measure how much or in what way the dependence relationships of two variables differ depending on their conditions. A canonical correlation may be used to quantify the relationship between two multivariate data, though it is still difficult to gauge the generalized dependence between two high-dimensional variables.

The proposed method works under the permutation null distribution and the testing procedure is based on the permutation test. The permutation test is easy to implement and avoids any parametric modeling assumptions. Moreover, it can ensure exact control of the type I error rate for all test statistics under the null hypothesis. These attractive properties have led to the permutation approach being used across a wide range of settings \cite{zhan2017fast, song2020generalized, liu2023kernel}. However, the permutation test has the requirement for exchangeability under the null hypothesis and the permutation approach could be problematic if the assumption of exchangeability is violated \cite{chung2013exact, chung2016multivariate, diciccio2017robust}. To address this issue, one could derive the asymptotic true null distribution of $\tilde{T}_{N}$ as it does not require the assumption of exchangeability. However, this is beyond the scope of the current endeavor and requires further investigation.  Nonetheless, our proposed approach represents a critical step towards more effective differential dependency analysis.


\bibliographystyle{IEEEtran}
\bibliography{Bibliography-MM-MC}


\end{document}